\documentclass[twocolumn,showpacs,preprintnumbers,amsmath,amssymb,pra]{revtex4}
\usepackage{amssymb}
\usepackage{amsmath}
\usepackage{graphics}
\newcommand{\bm}[1]{ \mbox{\boldmath $#1$}  }

\begin{document}

\title{Stability of a Fully Polarized Ultracold Fermi Gas near Zero-Crossing of a $p$-wave Feshbach Resonance}

\author{N.~T. Zinner}
\affiliation{ Department of Physics, Harvard University, Cambridge,
  Massachusetts 02138, USA}

\date{\today}

\begin{abstract}
We consider a fully polarized ultracold Fermi gas interacting through a $p$-wave Feshbach resonance. Using a two-channel
model, we find the effective potential at the point where the $p$-wave scattering length goes to zero. Here the
effective interaction provides attraction and one can therefore ask about the stability of the system. We calculate the energy
density of the system in the Thomas-Fermi approximation, determine the profile of the gas, and the critical number of particle
in the system as function of the relevant interaction parameters. The instability can be inferred from a simple breathing mode
argument which explains the scaling found numerically. The critical particle number turns out to be
extremely large unless the external confinement is very tight. We therefore conclude that the effect is insignificant 
for standard trapping potentials and
that the magnetic dipole interaction is the important term at zero scattering length. However, for tight confinement
as in an optical lattice higher-order corrections can become important.
\end{abstract}
\maketitle

\section{Introduction}
Feshbach resonances in ultracold atomic gases that allow tuning of interactions are
a continuing source of excitement. The use of $s$-wave resonances have led 
to great insight into f.x. the BCS-BEC crossover in the two-component Fermi gas \cite{bloch2008}.
Recently, also $p$-wave resonances have been explored in experiments with $^{40}$K \cite{regal03,ticknor04,gunter05,gaebler07} and $^{6}$Li \cite{zhang04,schunck05,chevy05}, and $p$-wave Feshbach molecules \cite{gubbels07} have been created \cite{gaebler07,fuchs08,inada08}.
This opens 
up the possibility of creating exotic $p$-wave superfluids for which several classical and quantum phase transitions 
have been predicted theoretically \cite{ho05,gurarie05,gurarie07,cheng05,cheng06,iskin06,pricoupenko06,jaksch02}.
Simultaneously, there has been a lot of recent interest in using Feshbach resonances to tune the scattering length to
zero \cite{roati2007,errico07,fattori2008a,fattori2008b,gustavsson2008,pollack2009}. These experiments work with Bose condensates and are able 
to probe residual magnetic dipole-dipole interactions at this zero-crossing \cite{fattori2008b}. However, as
discussed in \cite{zinner2009a,thoger2009,zinner2009c} there are also higher-order interaction terms 
present at zero-crossing that can influence the stability of the condensate for certain resonances.

In this work we want to explore 
the effects of higher-order interaction terms on Fermi gases as the $p$-wave scattering length
is tuned to zero. In particular, we study the fully polarized 
single-component system where $s$-wave interactions are Pauli suppressed and the $p$-wave term becomes important. Such highly polarized Fermi gases
are currently under intense study experimentally \cite{zwierlein06a,zwierlein06b,shin06,schunck07,partridge06a,partridge06b}. We will
study some bulk properties of the trapped system through the local-density approximation. We note that
the consequences of $p$-wave resonances have also recently been explored in traps for two- and three-particle systems \cite{yip08,jona08,idzi09}.

As we will show below, the higher-order correction term due to the $p$-wave Feshbach resonance is attractive and can therefore serve to 
destabilize the system when the particle number becomes large.
We consider the stability toward collapse within mean-field theory and find that 
a necessary condition for the higher-order terms to be important is that the external confinement be very tight, i.e. the trapping length scale
should be of order 0.03
$\mu$m or less. This is of course much smaller than typical traps, however, it might be relevant for optical lattice setups.

The paper is structured as follows. In Sec.~\ref{sec-pwave} we introduce and discuss a two-channel model 
for the $p$-wave resonance that is needed to describe the higher-order interaction term. We consider the full T-matrix at low momenta,
find the effective potential, and then match the parameters to low-energy $p$-wave scattering. Then in Sec.~\ref{sec-TF} we 
derive mean-field equations for the gas within the Thomas-Fermi or local-density approximation and discuss the density profiles.
In Sec.~\ref{sec-crit} we present some analytical stability conditions and calculate the critical number of particles
in the system as a function of the interaction strength. We compare and discuss experimentally relevant parameters in Sec.~\ref{sec-exp}. 
In the concluding Sec.~\ref{sec-conc} we summarize our findings and consider the relative importance of higher-order terms from the resonance to the magnetic dipole-dipole interaction. Here we again find that very tight confinement must be used for the higher-order correction to be of comparable magnitude.

\section{$p$-Wave Effective Potential}\label{sec-pwave}
Since we are interested in higher-order interaction terms at zero-crossing we use a
two-channel model to describe the $p$-wave Feshbach resonance. This model is obtained from
quantum defect theory \cite{mies84,julienne06}. If we assume that the energy of the bound
state varies linearly with magnetic field, $B$, then the $p$-wave phase-shift satisfies
\cite{idzi09}
\begin{align}
k^3\cot\delta_p(k)=-\frac{1}{f_{bg}(k)}\left[1-\frac{\Delta\mu\Delta B}{\Delta\mu(B-B_0)-\frac{\hbar^2k^2}{m}} \right]^{-1},
\end{align}
where $B_0$ is the resonance position, $\Delta B$ the width, $\Delta\mu$ the difference in magnetic moment
between the open and closed channel, and $m$ the reduced mass of the scattered atoms. The function $f_{bg}(k)$ represents the background scattering from the long-range
van der Waals (vdW) interaction. We will assume that $E\ll E_{vdW}$, where $E$ is the scattering energy and $E_{vdW}$ is the
energy scale of the vdW interaction. In this case we can ignore effective range corrections from the background term and 
we have $f_{bg}(k)\approx (a_{bg}^{p})^3$, where $a_{bg}^{p}$ is the background $p$-wave scattering length. 

We can now write the on-shell open-open $p$-wave T-matrix for the two-channel model at low energy as
\begin{align}
T_{p}(k)=&\frac{4\pi\hbar^2(a_{bg}^{p})^3}{m}&\nonumber\\
&\times\frac{k^2}
{\left(1-\frac{\Delta\mu\Delta B}{\Delta\mu(B-B_0)-\frac{\hbar^2k^2}{m}}\right)^{-1}+i(a_{bg}^{p}k)^3}.&
\end{align}
At this point one could attempt an effective-range expansion and
find the effective potential through the energy shift method \cite{roth01}. Unfortunately, at zero-crossing where
$\Delta B=B-B_0$ the coefficients of such an expansion are all divergent as discussed for the $s$-wave case in \cite{zinner2009c}.
However, since the full T-matrix is available we can follow the strategy of \cite{zinner2009c}. We go to zero-crossing
and expand the T-matrix to lowest order in $k$ to obtain
\begin{align}
T_{p}(k)=\frac{4\pi\hbar^2}{m}\frac{(a_{bg}^{p})^4 r_{e0}^{p}}{2} k^4+O(k^6),
\end{align}
where we have introduced the background effective range $r_{e0}^{p}=\tfrac{-2\hbar^2}{m\Delta\mu\Delta B a_{bg}^{p}}$.
Notice that $r_{e0}^{p}<0$ in this zero-range model \cite{phil98}. 
One can now solve the Lippmann-Schwinger equation to obtain the effective potential. Since we are interested in ultracold
fermions we take the long-wavelength limit, i.e. we let the cut-off momentum in the solution go to zero \cite{zinner2009c}. 
In that case we find the immediate answer $V=T$ and we merely have to transform the momentum-space expression into coordinate space. 
In our case we need a second-order $p$-wave zero-range term as outlined in \cite{roth01}. The properly symmetrized effective
potential is 
\begin{align}\label{veff}
V_{eff}^{p}=\frac{g_{p}^{(1)}}{2}\left[ 
\overleftarrow{\nabla}_{\bm r}^2\overleftarrow{\nabla}_{\bm r}\cdot\delta(\bm r)\overrightarrow{\nabla}_{\bm r}
  +\overleftarrow{\nabla}_{\bm r}\delta(\bm r)\cdot\overrightarrow{\nabla}_{\bm r}\overrightarrow{\nabla}_{\bm r}^2\right],
\end{align} 
where $g_{p}^{(1)}=\tfrac{4\pi\hbar^2}{m}\tfrac{(a_{bg}^{p})^4r_{e0}^{p}}{2}$ and $\bm r$ is the relative distance of the 
two atoms. The superscript on the coupling indicates that we consider the first correction term to the scattering length approximation \cite{roth01}.

Since $r_{e0}^{p}<0$ we have $g_{p}^{(1)}<0$. As we show in the next section, the negative coupling leads to an attractive interaction term in the energy density.
We can therefore ask about the stability of a single-component Fermi
gas interacting through a $p$-wave Feshbach resonance around zero-crossing. This is determined by the interplay of kinetic energy (or Fermi pressure) which tends to expand the gas and the attractive term above which causes contraction. If the latter is dominant we expect the system to collapse. 

The collapse proceed through recombination events into deeply bound dimer states in the van der Waals potential 
between the atoms. In the case of a fully polarized Fermi gas, $s$-wave collisions are suppressed due to the Pauli principle.
The leading order loss mechanism
is therefore in the $p$-wave channel \cite{suno03,levinsen08,jona08}. Close to a $p$-wave Feshbach resonance one can estimate the decay rate of the 
gas on both the BCS ($a_{p}<0$) and BEC ($a_{p}>0$) sides through the background parameters of the interaction as \cite{levinsen08} $\Gamma\sim \tfrac{\hbar}{m l^2}\tfrac{R_{vdW}}{l}$, where $R_{vdW}$ is the van der Waals length of the atoms \cite{pet02} and $l=n^{-1/3}$ is the interatomic distance when $n$ is the density of the gas. Based on this estimates one expects the lifetime of the polarized gases 
with a $p$-wave resonance to be much shorter than that of two-component gases with an $s$-wave resonance \cite{levinsen08}. In the current
work we do not consider the immediate vicinity of the resonance but rather zero-crossing where $a_{p}=0$. However, we expect the same
estimates to hold in this case as well. As we will see below, the collapse near zero-crossing is associated with an instability of the 
breathing mode which entails a change in the radial extension of the cloud. We thus imagine the collapse taking place through a decrease in 
cloud size and thus in $l$ with a corresponding increase in $\Gamma$. The instability should therefore manifest itself as a reduction of 
the lifetime of the system close to the critical number of particles calculated below. 

\section{Thomas-Fermi Energy Density}\label{sec-TF}
The energy density of a single-component homogeneous Fermi gas with the effective interaction of Eq.~\eqref{veff} can be
written in terms of the Fermi momentum, $k_F$. Here we use the Thomas-Fermi or local-density approximation to write 
the energy density in an external trapping potential in terms of the local Fermi momentum $k_{F}(\bm x)$ as
\begin{align}\label{edens}
\mathcal{E}_{TF}(\bm x)=&\frac{\hbar^2 k_{F}(\bm x)^{5}}{20\pi^2m}&\nonumber\\
&+\frac{1}{6\pi^2}V_{ext}(\bm x)k_{F}(\bm x)^{3}
+\frac{3g_{p}^{(1)} }{2800\pi^4 m}k_{F}(\bm x)^{10},&
\end{align} 
where the last term is obtained from the matrix elements of Eq.~\eqref{veff} in the Fermi gas 
(see \cite{roth01} and references therein for details). The local density 
is given by $\rho(\bm x)=k_F(\bm x)^3/6\pi^2$. The formalism we use is at zero temperature. This should be 
sufficient for the stability analysis as experiments have temperatures of $T/T_F\lesssim 0.05$ with
$T_F$ the Fermi temperature \cite{partridge06a,partridge06b}.
In the following we will assume that the 
external confinement has the form of an isotropic harmonic oscillator 
\begin{align}
V_{ext}(\bm x)=\frac{1}{2} \hbar\omega\left(\frac{\bm x}{b}\right)^2
\end{align}
with frequency $\omega$ and oscillator length $b$, although we note that isotropy is not essential
for the stability analysis presented below.

We now determine the ground-state density of the single-component Fermi system by minimizing the energy 
for a given number of particles. To this end we introduce the chemical potential, $\mu$, and perform the variation
\begin{align}
\frac{\delta}{\delta k_F(\bm x)}\left[\int d^3\bm x \left(\mathcal{E}_{TF}(\bm x)-\mu \rho(\bm x)\right)\right]=0
\end{align}
at each point $\bm x$. We then arrive at the equation
\begin{align}\label{eom}
\left[\frac{\mu}{\hbar\omega} -\frac{1}{2}\left(\frac{\bm x}{b}\right)^2\right]=&\frac{1}{2}(k_F(\bm x)b)^2&\nonumber\\
&+\frac{3}{70\pi}\left(\frac{a_{bg}^{p}}{b}\right)^4\frac{r_{e0}^{p}}{b}(k_F(\bm x)b)^7,&
\end{align}
where explicit oscillator units have been introduced. For each $\bm x$ this represents a polynomial equation which
can be solved for $k_F(\bm x)$. The particle number, $N$, can be adjusted by varying $\mu$ to ensure that $N=\int d^3\bm x \rho(\bm x)$.

\begin{figure}
\centering
\resizebox{0.9\columnwidth}{!}{%
  \includegraphics{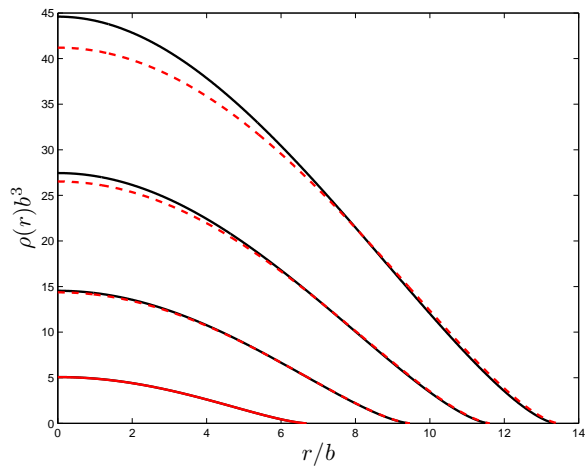}
}
  \caption{Spherically symmetric density profiles $\rho(r)$ for $\mu/\mu_{max}=0.25$, 0.5, 0.75, and 1.0 (from bottom to top) where $\mu_{max}$ is the critical chemical potential calculated for the $^{40}$K resonance discussed in the text and trap size $b=0.1\mu$m. The solid (black) lines are obtained from Eq.~\eqref{eom}, whereas the dashed (red) lines are the non-interacting densities for the same particle number. The particle numbers are $N=1.9\cdot 10^3$, $1.5\cdot 10^4$, $5.1\cdot 10^4$, and $1.2\cdot 10^5$ (bottom to top).}
\label{fig-dens}      
\end{figure}

In Fig.~(\ref{fig-dens}) we show some typical density profiles obtained by solving Eq.~\eqref{eom} for $\mu/\mu_{max}=0.25$, 0.5, 0.75, and 1.0 (from bottom to top), where $\mu_{max}$ is the critical chemical potential discussed in the next section. The full (black) lines are obtained 
from Eq.~\eqref{eom}, whereas the dashed (red) lines are the corresponding non-interacting systems in the trap with the chemical potential
adjusted to yield the same particle number. The interacting densities are in general higher in the center of the trap as expected from the attractive nature of the interaction at zero-crossing. In fact, close to the critical number of particles we see a clear difference in the density profiles, whereas for the lower particle numbers the interacting and non-interacting densities are very similar. The effects of higher-order interactions around zero-crossing can thus in principle be probed through the density profile of the gas. In the momentum distribution there should also be a signature of the squeezing of the cloud due to the attraction.

\subsection{Critical Particle Number}\label{sec-crit}
From the structure of Eq.~\eqref{eom} and the fact that $g_{p}^{(1)}<0$ we see that there will be a maximum Fermi momentum, $k_{max}$, and a corresponding $\mu_{max}$ beyond which no solution can be found for a given $\bm x$ (as discussed for the lowest order $p$-wave interaction in \cite{roth01}). This means that there will be a critical number of particles, $N_{max}$, so that for $N>N_{max}$ the system is unstable in mean-field theory. We now determine these quantities.

In the center of the harmonic trap we have $V_{ext}(\bm 0)=0$, and away from the center $\mu/\hbar\omega-(x/b)^2/2$ is monotonically decreasing toward the edge at $x/b=\sqrt{2\mu/\hbar\omega}$. The maximum Fermi momentum $k_{max}$ is therefore obtained by finding the turning point of the right-hand side of Eq.~\eqref{eom}. This yields
\begin{align}\label{kmax}
k_{max}b=\left[\frac{10\pi}{3\alpha}\right]^{1/5}, \quad \mu_{max}=\frac{5}{14}\hbar\omega \left(k_{max}b\right)^{2},
\end{align}
where we have introduced $\alpha=(a_{bg}^{p}/b)^4|r_{e0}^{p}/b|$ to reduce the notation. In terms of density, the stability condition is $\alpha^{3/5}\rho(\bm x) b^3\leq 0.07$, which must hold everywhere in the trap for the system to remain stable. As discussed in \cite{roth01}, this condition is independent of the trap geometry and the use of an isotropic harmonic trap is therefore not essential for the stability analysis.

\begin{figure}
\centering
\resizebox{0.9\columnwidth}{!}{
  \includegraphics{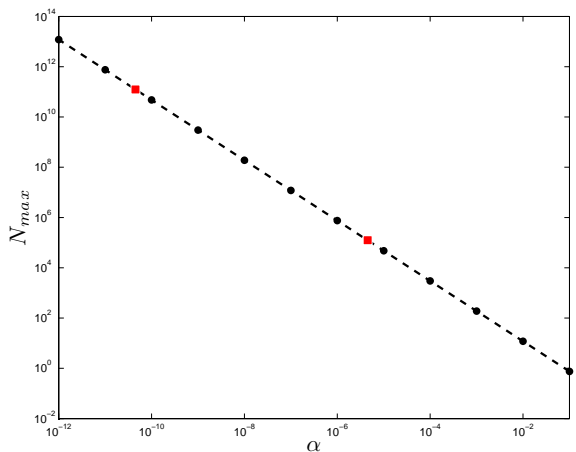}
}
  \caption{Critical particle number, $N_{max}$, as function of the interaction strength parametrized by $(a_{bg}^{p}/b)^4|r_{e0}^{p}/b|$. The (black) circles indicate the decades. The (red) squares are for the $^{40}$K resonance discussed in 
  the text with oscillator length $b=1\mu$m (left) and $b=0.1\mu$m (right).}
  \label{fig-crit}
\end{figure}

From the values of $k_{max}$ above we can calculate the density profile for $\mu=\mu_{max}$ and determine $N_{max}$ by simple integration. In Fig.~(\ref{fig-crit}) we plot $N_{max}$ as a function of the interaction strength parametrized by $\alpha$ on a log-log scale. We indeed see that the critical number of particles grows rapidly with decreasing interaction strength. The (red) squares denote some estimates for a realistic $p$-wave resonance in $^{40}$K (to be discussed below) with the upper one corresponding to $b=1\mu$m and the lower one having $b=0.1\mu$m. In the former case we find $N_{max}\sim 10^{11}$ whereas the tighter trapped system has $N_{max}\sim 10^5$ which is in the regime of experiments with polarized Fermi gases \cite{zwierlein06a,zwierlein06b,shin06,schunck07,partridge06a,partridge06b}. From the slope in Fig.~(\ref{fig-crit}) we find $N_{max}\propto\alpha^{-1.2}$. Decreasing the trap length by one order of magnitude is therefore seen to decrease $N_{max}$ by six orders of magnitude (in accord with the difference between the (red) squares in Fig.~(\ref{fig-crit})).

We now proceed to show how the scaling of $N_{max}$ with $\alpha$ is related to the instability of the monopole or breathing mode of the system. To see this we employ the method of collective coordinates \cite{pet02}, although the same result can be found through the sum-rule approach \cite{vichi99}. We assume that the cloud maintains its shape but allow for variations of the spatial extent parametrized by the radius $R$. Considering small deviations from equilibrium we can derive a virial theorem of the form
\begin{align}
-2E_{K}+2E_{V}-7E_{I}=0,
\end{align} 
where the kinetic, $E_{K}$, harmonic trapping, $E_{V}$, and interaction, $E_{I}$, parts are the integrated equivalents of the terms in Eq.~\eqref{edens}. With this relation, we can relate the frequency of the breathing mode, $\omega_B$, to the energies through
\begin{align}
\omega_{B}^{2}=\omega^{2}\left[4+\frac{35}{2}\frac{E_I}{E_V}\right].
\end{align}
The instability occurs since $E_I\propto -\alpha<0$. For sufficiently large $\alpha$ the breathing mode frequency will therefore become imaginary indicating an instability toward collapse when $E_I/E_V<-8/35$.

The dependence of the instability condition on $\alpha$ and $N$ can be found using the simple non-interacting density $k_F(\bm x)=\sqrt{2\mu/\hbar\omega-{\bm x}^2/b^2}/b$. We find $E_V=c_{V}N^{4/3}$ and $E_I=-\alpha c_{I}N^{13/6}$, where $c_V$ and $c_I$ are constants. The instability condition is therefore $\alpha N_{max}^{5/6}=c$, with $c$ a constant of order one. We thus recover the scaling $N_{max}\propto \alpha^{-1.2}$ as found numerically from the interacting densities above.

\section{Comparison to Experiments}\label{sec-exp}
We now estimate the parameters needed in our scenario for $p$-wave Feshbach resonances in $^{40}$K \cite{ticknor04}. These data have been fitted to the $p$-wave resonance model introduced above for all magnetic substates $m_l=0,\pm 1$ of the $|f=\tfrac{9}{2},m_f=-\tfrac{7}{2}\rangle$ hyperfine state in \cite{idzi09}. The parameters needed here are similar for all $m_l$ and for concreteness we focus on the $m_l=0$ state. The resonance is found at $B_0=198.85$G with a width of $\Delta B=-20.342$G, a background scattering length of $a_{bg}^{p}=-104.26a_0$ (with $a_0$ the Bohr radius), and magnetic moment difference $\Delta\mu=0.067\mu_B$ where $\mu_B$ is the Bohr magneton. For these values we find $r_{e0}^{p}=-914.72a_0$. 
From $\Delta B$ alone the resonance appears rather broad and a small value of the effective range would be expected. However, the small value of $\Delta\mu$ compensates and increases $r_{e0}^{p}$. In fact, $p$-wave resonances are in general very narrow as discussed in \cite{gubbels07}.

For the interaction parameter introduced above we have 
\begin{align}
\alpha=
1.08\cdot 10^{11} (\tfrac{a_0}{b})^5=4.48\cdot10^{-11}\left(\frac{1\,\mu\text{m}}{b}\right)^5,
\end{align}
and we thus immediately see that the tightness of the trapping is the decisive factor. Comparing to Fig.~(\ref{fig-crit}) we
see that typical experimental trap lengths of $b=1\mu$m will yield critical particle numbers that are several orders of magnitude 
above current experimental numbers (upper (red) square). However, as mentioned above, an order of magnitude decrease in the trap 
length will bring the number into the experimental range (lower (red) square). The density profiles for $b=0.1\mu$m are shown in
Fig.~(\ref{fig-dens}) together with the non-interacting densities and at the critical number we do see considerable deviations within
the mean-field Thomas-Fermi picture. We conclude that tight trapping with $b\lesssim 0.1\mu$m is necessary in order to experimentally 
observe any effect of higher-order interactions on the stability of the system around the zero-crossing of realistic $p$-wave Feshbach resonances.

In this work we have exclusively employed the zero temperature formalism. One can of course extend the semi-classical approach to 
finite temperature to get a better description of experiments. Here we will be content with qualitative estimates of the 
effect of temperature which can be obtained by considering the non-interacting Fermi gas in a 
harmonic trap. As shown in the earlier work of \cite{butts97}, the intuitively clear effect of finite temperature is to 
extend the density profile of the gas (see Fig.~(3) of \cite{butts97}). In turn, the local Fermi wavevector is reduced and 
likewise the effect of interaction in our Eq.~\eqref{edens}. We therefore expect the critical number to increase with temperature. 
However, as noted above experiments can cool to only 5-10\% of the Fermi temperature and for this level of 
degeneracy the deviation from the $T=0$ density profile should be very small \cite{butts97}. This justifies the 
use of the zero temperature formalism.

\section{Conclusions}\label{sec-conc}
We have studied the effects of higher-order interactions in single-component Fermi gases interacting via a $p$-wave
Feshbach resonance as the scattering length goes to zero. Using the low-energy limit of the full T-matrix we found a formula for 
the effective potential with strength proportional to the background parameters of the resonance. Implementing this effective interaction in a
local-density Thomas-Fermi approximation for the Fermi gas in an isotropic harmonic oscillator trap, we found an equation for the 
density profile. Since the effective interaction around zero-crossing is attractive, the system has a critical particle number beyond which it becomes unstable according to mean-field theory. We gave an analytic expression for the stability condition and numerically calculated the 
critical particle number for relevant experimental parameters. The density profile was compared to the non-interacting profile and 
considerable deviations found near the critical particle number. In comparison to experimental parameters, we find that the external potential must provide a very tight confinement with trap length below 0.1$\mu$m for the critical number to be at the value of experiments.

At zero-crossing there are other residual forces in the gas. In particular, as was recently found in Bose gases \cite{fattori2008a,fattori2008b}, the magnetic dipole-dipole interaction must be taken into account. We therefore have to consider the
relative strength of this interaction to the higher-order $p$-wave term discussed in the current work. Comparing the couplings
we find \cite{pet02}
\begin{align}
\left|\frac{U_{md}}{U_{g_{p}^{(1)}}}\right|\sim &\frac{a_0 b^4}{\left(a_{bg}^{p}\right)^4 r_{e0}^{p}}=&\nonumber\\
&1.3\cdot10^{6}\left[\frac{100a_0}{a_{bg}^{p}}\right]^4\left(\frac{1000a_0}{|r_{e0}^{p}|}\right)\left[\frac{b}{1\,\mu\text{m}}\right]^4,&
\end{align}
where the trap length scale $b$ appears since the effective potential carries four
derivatives. For the ratio to be of order one, a trapping length of $b\sim 0.03\mu$m is needed. With normal traps the magnetic dipole-dipole interaction is thus much stronger at zero-crossing. Curiously, we note that with an optical lattice of some 30 lattice sites in a beam waist of $1\mu$m the lattice site size is roughly $0.03\mu$m. Exploring the effects of higher-order interactions in optical lattices is therefore a promising direction in which the current work can be extended. Of course, this also holds true for higher-order effects near zero-crossings for $s$-wave Feshbach resonances and work in this direction is also in progress.

The attractive nature of the higher-order correction could perhaps also lead to $p$-wave superfluidity at 
zero-crossing. We can make an estimate of the critical temperature for this phase if we assume the same 
functional form as for $s$-wave superfluidity in two-component gases. As we expect it to 
increase with interaction strength and with Fermi wavevector, $T_c\sim T_F\exp\left(-\tfrac{1}{\alpha (k_F b)^5}\right)$.
In the center of the trap, $k_F=k_{max}$, and from Eq.~\eqref{kmax} we find $T_c\sim T_F \exp(-3/10\pi)\approx 0.9T_F$ which 
is a large value compared to experimentally realizable temperatures. However, factors of order 1-10 both inside and 
outside the exponential can severely reduce $T_c$ as seen in the $s$-wave case \cite{pet02}, and we also have to keep 
in mind that $k_F$ decreases toward the edge of the trap. A potentially much more severe hindrance is the fact that 
the estimates of the decay rate discussed above also suggest that the $p$-wave superfluid does not have time to form before 
it decays \cite{levinsen08}.

The above estimate demonstrates that the dipole-dipole interaction is overwhelmingly dominant in large
traps, and an optical lattices is needed for the correction to matter. Transition temperatures could be raised by manipulating the lattice symmetries as found in \cite{iskin05} and a stabilizing mechanism in the lattice was also recently proposed \cite{han09}. 
We aim to explore the
influence of the higher-order corrections on such scenarios as well. We caution, however, that despite detailed experimental studies of $p$-wave Feshbach resonances \cite{regal03,ticknor04}, an important parameter related to the Feshbach channel coupling cannot yet be deduced from the data and the possibility of producing $p$-wave superfluids is therefore still hard to estimate \cite{gurarie07}. In contrast the $p$-wave resonance model used above only requires background parameters and the magnetic moment difference. The evaluation of instability in mean-field theory is therefore a simpler problem. We still have to remember that our parameters where obtained from a fitting procedure of the data in \cite{ticknor04} to a two-channel model which is accurate to within a few percent \cite{idzi09}.

\paragraph*{Acknowledgements}
The author thanks David Pekker and Chris Pethick for discussions and comments. This work was supported by the Villum Kann Rasmussen foundation.

\end{document}